# Conformality Emergent In String Phenomenology

Paul H. Frampton

Department of Physics and Astronomy, University of North Carolina, Chapel Hill, NC 27599-3255, USA.

**Abstract.** Use of the AdS/CFT correspondence to arrive at phenomenological gauge field theories is discussed, focusing on the orbifolded case without supersymmetry. An abelian orbifold with the finite group Z_p can give rise to a G = U(N)^p gauge group with chiral fermions and complex scalars in different bi-fundamental representations of G. The naturalness issue is discussed, particularly the absence of quadratic divergences in the scalar propagator at one loop. This requires that the scalars all be in bi-fundamentals with no adjoints, coincident with the necessary and sufficient condition for presence of chiral fermions. Speculations are made concerning new gauge and matter particles expected soon to be pursued experimentally at the LHC.

## QUIVER GAUGE THEORY

The relationship of the Type IIB superstring to conformal gauge theory in d=4 gives rise to an interesting class of gauge theories. Choosing the simplest compactification [1] on AdS_5 X S_5 gives rise to an **N** = 4 SU(N) gauge theory which is known to be conformal due to the extended global supersymmetry and non-renormalization theorems. All of the RGE $\beta$-functions for this **N** = 4 case are vanishing in perturbation theory. It is possible to break the **N**=4 to **N**=2, 1, 0 by replacing S_5 by an orbifold S_5/$\Gamma$ where $\Gamma$ is a discrete group with $\Gamma \subset$ SU(2) $\subset$SU(3), $\not\subset$ SU(3) respectively. In building a conformal gauge theory model [2,3,4], the steps are: (1) Choose the discrete group $\Gamma$ ; (2) Embed $\Gamma \subset$ SU(4); (3) Choose the N of SU(N); and(4) Embed the Standard Model SU(3) X SU(2) X U(1) in the resultant gauge group U(N)^p (quiver node identification). Here we shall look only at abelian $\Gamma$ = Z_p and define $\alpha$ = exp(2 $\pi$ i/p). It is expected from the string-field duality that the resultant field theory is conformal in the N $\rightarrow \infty$ limit, and will have a fixed manifold, or at least a fixed point, for N finite. Before focusing on the non-supersymmetric cases, let us first examine an **N**=1 model put forward in [5].The choice is $\Gamma$ = Z_3 and the **4** of SU(4) is **4** = (1, $\alpha$ , $\alpha$ , $\alpha$ ^2). Choosing N=3 this leads to the three chiral families under SU(3)^3 trinification [6]

$$3 [(3, 3^*, 1)+ (1, 3, 3^*) + (3^*, 1, 3)]$$

In this model it is interesting that the number of families arises as 4-1=3, the difference between the 4 of SU(4) and the number of unbroken supersymmetries. However this model has no gauge coupling unification; also, keeping supersymmetry is against the spirit of the conformality approach. We now address examples which accommodate three chiral families, break all supersymmetries and possess gauge coupling unification, including the correct value of the electroweak mixing angle.

## GAUGE COUPLINGS

An alternative to conformality, grand unification with supersymmetry, leads to an impressively accurate gauge coupling unification [7]. In particular it predicts an electroweak mixing angle at the Z-pole, $\sin^2 \theta = 0.231$. This result may, however, be fortuitous, but rather than abandon gauge coupling unification, we can re-derive $\sin^2 \theta = 0.231$ in a different way by embedding the electroweak SU(2) X U(1) in U(N) X U(N) X U(N) to find $\sin^2 \theta = 3/13 \cong 0.231$ [8]. This will be a common feature of the models in this presentation.

For the conformal theories to be finite without quadratic or logarithmic divergences. This requires appropriate equal number of fermions and bosons which can cancel in loops and which occur without the necessity of space-time supersymmetry. As we shall see in one example, it is possible to combine space-time supersymmetry with conformality but the latter is the driving principle and the former is merely an option: additional fermions and scalars are predicted by conformality in the TeV range [8], but in general these particles are different and distinguishable from supersymmetric partners. The boson-fermion cancellation is essential for the cancellation of infinities, and will play a central role in the calculation of the cosmological constant (not discussed here).

What is needed first for the conformal approach is a simple model. Here we shall focus on abelian orbifolds characterized by the discrete group $Z_p$. Non-abelian orbifolds have been systematically analyzed elsewhere [9].The steps in building a model for the abelian case (parallel steps hold for non-abelian orbifolds) are:

**I**. Choose the discrete group $\Gamma$. Here we are considering only $\Gamma = Z_p$. We define $\alpha = \exp(2\pi i/p)$.

**II**. Choose the embedding of $\Gamma \subset SU(4)$ by assigning $\mathbf{4} = (\alpha^{A_1}, \alpha^{A_2}, \alpha^{A_3}, \alpha^{A_4})$ such that $\sum A_q = 0$ (mod p). To break $\mathbf{N} = 4$ supersymmetry to $\mathbf{N} = 0$ (or $\mathbf{N} = 1$) requires that none (or one) of the $A_q$ is equal to zero (mod p).

**III.** For chiral fermions one requires that $\mathbf{4} \neq \mathbf{4^*}$ for the embedding of $\Gamma$ in SU(4). The chiral fermions are in the bifundamental representations of $U(N)^p$

$$\sum (N_i, \bar{N}_{i + A_q})$$

If $A_q=0$ we interpret $(N_i, \bar{N}_i)$ as a singlet plus an adjoint of $SU(N)_i$.

**IV**.The **6** of SU(4) is real $\mathbf{6} = (a_1, a_2, a_3, -a_1, -a_2, -a_3)$ with $a_1 = A_1 + A_2$, $a_2 = A_2 + A_3$, and $a_3 = A_3 + A_1$ (recall that all components are defined modulo p).

The complex scalars are in the bi-fundamentals

$$\sum (N_i, \bar{N}_{i \pm a_j})$$

The condition in terms of $a_j$ for $\mathbf{N} = 0$ is $\sum (\pm a_j) \neq 0 \pmod{p}$.

**V.** Choose the N of U(Nd_i) (where the d_i are the dimensions of the representations of $\Gamma$ ). For the abelian case where d_i = 1, it is natural to choose N=3 the largest SU(N) of the standard model (SM) gauge group. For a non-abelian $\Gamma$ with d_i ≠ 1 the choice N=2 would be indicated.

**VI.** The p quiver nodes are identified as color (C), weak isospin (W) or a third SU(3) (H). This specifies the embedding of the gauge group $SU(3)_C \times SU(3)_W \times SU(3)_H \subset U(N)^p$. This quiver node identification is guided by (7), (8) and (9) below.

**VII.** The quiver node identification is required to give three chiral families. It is sufficient to make three of the (C + A_q) to be W and the fourth H, given that there is only one C quiver node, so that there are three (3, 3*, 1). Provided that (3*, 3, 1) is avoided by the (C - A_q) being H, the remainder of the three family trinification will be automatic by chiral anomaly cancellation. Actually, a sufficient condition for three families has been given; it is necessary only that the difference between the number of (3 + A_q) nodes and the number of (3 - A_q) nodes which are W is equal to three. We assume no fourth family [10].

**VIII.** The complex scalars of must be sufficient for their vacuum expectation values (VEVs) to spontaneously break $U(3)^p \rightarrow SU(3)_C \times SU(3)_W \times SU(3)_H \rightarrow SU(3)_C \times SU(2)_W \times U(1)_Y \rightarrow SU(3)_C \times U(1)_Q$. Note that, unlike grand unified theories (GUTs) with or without supersymmetry, the Higgs scalars are here prescribed by the conformality condition. This is more satisfactory because it implies that the Higgs sector cannot be chosen arbitrarily, but it does make model building more interesting.

**IX.** Gauge coupling unification should apply at least to the electroweak mixing angle $\sin^2\theta = g_Y^2 / (g_2^2 + g_Y^2) \cong 0.231$. For trinification $Y = 3^{-1/2}(-\lambda_{8W} + 2\lambda_{8H})$ so that $(3/5)^{1/2}$ is correctly normalized. If we make $g_Y^2 = (3/5)g_1^2$ and $g_2^2 = 2 g_1^2$ then $\sin^2\theta = 3/13 \cong 0.231$ with sufficient accuracy.

## 4 TEV GRAND UNIFICATION

This topic which explains the origin of the scale 4 TeV is omitted for reasons of time; details are in [11].

## NATURALNESS REFERENCES

Hierarchy and naturalness were concepts introduced in the late 1960s, particularly by Wilson. Some of it is written in [12]. Three decades later in December 2004, there is a contrary viewpoint in [13] where the author calls his earlier objection to the occurrence of scalar fields in quantum field theory a "blunder".

Other references are worth mentioning. In this approach, 333-trinification emerges more readily than 422-Pati-Salam and interesting work was done recently by this symposium's Chair [14]. Triangle anomalies will play a role in the sequel and a seminal paper is [15]. The conformaility approach is related in general terms to the misaligned supersymmetry proposed in [16].

Superconformal symmetry illustrated by the remarkable **N** = 4 SU(N) Yang-Mills theory, and discovered [17] in 1983 has probably a lot to do with extending he standard model. This theory was connected to string theory, for infinite N, in [1] by the AdS/CFT correspondence. For phenomenology the symmetry must be lessened, maybe to supersymmetry whose proposal in 1974 answered Wilson's objection about quadratic divergences in the scalar propagator. Alternatively, it could be lessened to conformality, meaning four-dimensional conformal invariance at high energy above 4 TeV as may be obtained through the reduction of **N** = 4 $\rightarrow$ **N** = 0 by orbifolding.

It is an interesting sociological comment that there are 10,000 papers on supersymmetry but less than 100 on conformality. This is reminiscent of one review [18] of superstrings which commented on the low number of papers.

# ABELIAN QUIVERS

Classification of abelian quiver gauge theories is a useful step. We consider the compactification of the type-IIB superstring on the orbifold $AdS_5 \times S^5/\Gamma$ where $\Gamma$ is an abelian group $\Gamma = Z_p$ of order p with elements $\exp(2\pi i A/p)$, $0 < A < (p-1)$. The resultant quiver gauge theory has **N** residual supersymmetries with **N** = 2,1,0 depending on the details of the embedding of $\Gamma$ in the SU(4) group which is the isotropy of the $S^5$. This embedding is specified by the four integers $A_m$, $1 < m < 4$ with

$$\sum A_m = 0 \pmod{p}$$

which characterize the transformation of the components of the defining representation of SU(4). We are here interested in the non-supersymmetric case **N** = 0 which occurs if and only if all four $A_m$ are non-vanishing.

The gauge group is $U(N)^p$. The fermions are all in the bi-fundamental representations

$$\sum\sum (N_j, \bar{N}_{j + A_m})$$

which are manifestly non-supersymmetric because no fermions are in adjoint representations of the gauge group. Scalars appear in representations

$$\sum\sum (N_j, \bar{N}_{j \pm a_i})$$

in which the six integers $(a_i, -a_i)$ characterize the transformation of the antisymmetric second-rank tensor representation of SU(4). The $a_i$ are given by $a_1 = (A_2+A_3)$, $a_2 = (A_3+A_1)$, $a_3 = (A_1+A_2)$. It is possible for one or more of the $a_i$ to vanish in which case the corresponding scalar representation in the summation is to be interpreted as an adjoint representation of one particular $U(N)_j$. One may therefore have zero, two, four or all six of the scalar representations in such adjoints.

For the lowest few orders of the group $\Gamma$, the members of the infinite class of N=0 abelian quiver gauge theories can be tabulated [19] and one finds that the scalars can be in bi-fundamentals and/or adjoints.

To be of more phenomenolgical interest the model should contain chiral fermions. This requires that the embedding be complex: $A_m$ not equivalent to $-A_m$ (mod p). It will now be shown that for the presence of chiral fermions all scalars must be in bifundamentals.

The proof of this assertion follows by assuming the contrary, that there is at least one adjoint arising from, say, $a_1=0$. Therefore $A_3=-A_2$ (mod p). But then it follows that $A_1=-A_4$ (mod p). The fundamental representation of SU(4) is thus real and fermions are non-chiral. (This is almost obvious but for a complete justification, see [20]).

It follows that: <u>In an N=0 quiver gauge theory, chiral fermions are present if and only if all scalars are in bi-fundamental representations.</u>

# QUADRATIC DIVERGENCES

The lagrangian for the nonsupersymmetric $Z_p$ theory is written in a convenient notation which accommodates simultaneously both adjoint and bifundamental scalars in *e.g.*[19].

As we showed in the previous section, the infinite sequence of nonsupersymmetric $Z_p$ models can have scalars in adjoints (corresponding to $a_i = 0$) and bifundamentals ($a_i$ not equal to 0). Denoting by x the number of the three $a_i$ which are non-zero, the models with x=3 have only bifundamental scalars, those with x=0 have only adjoints while x=1,2 models contain both types of scalar representations.

As we have seen, to contain the phenomenologically-desirable chiral fermions, it is necessary and sufficient that x=3. Let us first consider the quadratic divergence question in the mother **N** = 4 theory. The **N**=4 lagrangian is like that for the **N**=0 quiver but since there is only one node all those subscripts become unnecessary so the form is simpler, see [19]. All **N** = 4 scalars are in adjoints and the scalar propagator has one-loop quadratic divergences coming potentially from three scalar self-energy diagrams:

(a) the gauge loop (one quartic vertex);

(b) the fermion loop (two trilinear vertices);

and (c) the scalar loop (one quartic vertex).

For **N** = 4 the respective contributions of (a, b, c) are computed to be proportional to (1, -4, 3) which cancel exactly.

The **N** = 0 results for the scalar self-energies (a, b, c) are computable from the lagrangian of **N**=0. The result is pleasing. The quadratic divergences cancel if and only if x = 3, exactly the same ``if and only if'' as to have chiral fermions.

It is pleasing that one can independently confirm the results directly from the interactions in the **N**=0 lagrangian. To give just one explicit example, in the contributions to diagram (c) from the last term in the **N**=0 lagrangian [19], the 1/N corrections arise from a contraction of $\Phi$ with $\Phi$ when all the four color superscripts are distinct and there is consequently no sum over color in the loop. For this case, examination of the node subscripts then confirms proportionality to the Kronecker delta, $\delta\{0, a_i\}$. If and only if all $a_i$ are not equal to 0, all the other terms in N=0 do not lead to 1/N corrections to the N=4.

Some comments on the literature are necessary. In one paper [21] it was claimed that there are always $1/N$ corrections to spoil cancellation for finite N and that N > $10^{28}$ is necessary! This was because of a technical error that the orbifolded gauge group is not $SU(N)^p$ but $U(N)^p$ and bifundamentals carry U(1) charges. A paper by Fuchs [22] which partially corrected this point.

The conclusion is that the chiral $Z_7$, $Z_{12}$ models proposed respectively in [8, 11] which contain the standard model are free of one-loop quadratic divergences in the scalar propagator.

Nevertheless the overall conformal invariance would not be respected by U(1) factors which would have non-zero positive beta-functions, unless additional contributions are needed also for anomaly cancellation. A better understanding of these U(1)'s may be necessary to achieve the hope of a fully four-dimensionally conformally invariant extension of the standard model. Eventually gravity, at the Planck scale, will inevitably break conformal invariance because Newton's constant is dimensionful. A realistic hope is that there is a substantial window of energy scales where conformal invariance is an excellent approximation between, say, 4 TeV for at least a few orders of magnitude in energy even towards a scale approaching the see-saw scale of about $10^{10}$ GeV. It is difficult to foresee how large the conformality window is.

Finally it is interesting to note that the present models seem to have all the ingredients of the so-called little Higgs models, which were proposed later [23], with the quiver diagram here interpreted as the theory space there.

This conformality idea, that an augmented standard model possess an energy window of conformal invariance starting just above the weak interaction scale, requires the existence of new undiscovered particles accessible to the LHC: gauge bosons which fill out the unitary gauge group $U(N)^p$ which contains the established SU(3) X SU(2) X U(1); chiral fermions in bifundamental representations of U(N)^p; and, as shown in the present article, complex scalars also in bifundamentals of U(N)^p. The new experimental results should be able to distinguish these definite predictions coming from the assumption of four-dimensional conformal invariance.

## CONFORMALITY SUMMARY

**I.** Nonsupersymmetric quiver gauge theories motivated by AdS/CFT correspondence are very interesting to model builders.

**II.** Phenomenology of conformality has striking resonances with the standard model.

**III.** 4 TeV Unification predicts three families and new particles around 4 TeV accessible to experiment (LHC) [see 11 for details].

**IV.** The scalar propagator in these theories has no quadratic divergence if and only if there are chiral fermions precisely as required in the standard model.

## EXPERIMENTAL EVIDENCE

Two pieces of experimental evidence for conformality are:

**I.** The representation content of the 321-standard model can all be accommodated by bi-fundamentals since *e.g.* (8,2) and (3,3) of SU(3)XSU(2) are not present in Nature.

**II.** The electric neutrality of the H atom supports accommodation of the U(1) in a non-abelian SU(3) as specified by conformality.

## WHAT WILL LHC FIND?

Each particle phenomenologist, especially if active in the 1970s, is asked: what will the LHC discover? Since space permits, here are my wishes:

**I.** There will be one Higgs boson with mass somewhere between 114 GeV and 300 GeV. It will appear to be an elementary scalar at LHC.

**II.** There will be additional gauge bosons in the TeV region signaling extension of the established standard 321 gauge group (321 = SU(3) X SU(2) X U(1)) to 333 (possibly more 3's). I think this more likely than 422, 3211, 32111, etc.

**III.** There will be additional quarks and leptons, maybe scalars, in the TeV region.

The particles discovered in **III** will lead to confirmation of conformality [2]. The gauge bosons discovered in **II** will hopefully confirm both the axigluon [24] and bilepton [25].

## ACKNOWLEDGEMENTS

This work was supported in part by the Office of High Energy, US Department of Energy under Grant No. DE-FG02-97ER41036.